\documentclass{article}

\usepackage{microtype}
\usepackage{graphicx}
\usepackage{booktabs} 

\usepackage[accepted]{icml2019}

\usepackage{amssymb}
\usepackage{amsmath}
\usepackage[latin1]{inputenc}
\usepackage{mathabx}
\usepackage{pifont}
\usepackage{setspace}
\usepackage{psfrag}
\usepackage{color}
\usepackage{multirow}
\usepackage{wasysym}
\usepackage{times,balance,amsfonts,pgf,tikz,pgffor}
\usepackage{pgfplots}
\usetikzlibrary{shadows.blur}
\usetikzlibrary{shapes.symbols}
\usepackage[short]{optidef}
\usepackage[font=small,labelfont=bf]{caption}

\usepackage[bf]{subfigure}
 \usetikzlibrary{positioning}
\usepackage{tabularx}
\usepackage{algorithm}
\usepackage[noend]{algpseudocode}
\usetikzlibrary{patterns}

\makeatletter
\def\algbackskip{\hskip-\ALG@thistlm}
\makeatother

\makeatletter
\def\BState{\State\hskip-\ALG@thistlm}
\makeatother

\makeatletter
\newcommand{\multiline}[1]{%
	\begin{tabularx}{\dimexpr\linewidth-\ALG@thistlm}[t]{@{}X@{}}
		#1
	\end{tabularx}
}
\makeatother

\newcommand{\idxset}{\mathcal{I}}

\newcommand{\fMAC}{g}

\newcommand{\ie}{i.e., }

\newcommand{\taskbox}{$\text{task block }$}
\newcommand{\taskboxes}{$\text{task blocks }$}

\newcommand{\MAC}{$\text{basic }$}

\newcommand{\infoblock}{$\text{information block }$}

\newcommand{\infoblocks}{$\text{information blocks }$}
\newcommand{\infoblocksno}{$\text{information blocks}$}

\newcommand{\taskmtxs}{$\text{matrices }$}

\newcommand{\layer}{$\text{layer }$}

\newcommand{\layers}{$\text{layers }$}

\newcommand{\datachunk}{$\text{submatrix }$}
\newcommand{\datachunks}{$\text{submatrices }$}

\newcommand{\datachunksno}{$\text{submatrices}$}

\newcommand{\encdatachunks}{encoded submatrices }

\newcommand{\tcomp}{T}

\newcommand{\floor}[1]{\lfloor{#1}\rfloor}
\newcommand{\abs}[1]{\left|#1\right|}

\newcommand{\seti}[1]{\mathcal{S}_{#1}}

\newcommand{\setxi}[1]{\mathcal{S}_{x#1}}
\newcommand{\setyi}[1]{\mathcal{S}_{y#1}}
\newcommand{\setzi}[1]{\mathcal{S}_{z#1}}
\newcommand{\setxij}[2]{\mathcal{S}_{x#1,#2}}
\newcommand{\setyij}[2]{\mathcal{S}_{y#1,#2}}
\newcommand{\setzij}[2]{\mathcal{S}_{z#1,#2}}

\newcommand{\inR}[3]{#1 \in \mathbb R^{#2 \times #3}}

\newcommand{\matA}{\mathbf{A}}  
\newcommand{\matB}{\mathbf{B}}

\newcommand{\eA}[2]{\textit a_{#1,#2}}
\newcommand{\eB}[2]{\textit b_{#1,#2}}

\newcommand{\Xdir}{x}
\newcommand{\Ydir}{y}
\newcommand{\Zdir}{z}
\newcommand{\divx}{M_{x\ly}}
\newcommand{\divy}{M_{y\ly}}
\newcommand{\divz}{M_{z\ly}}

\newcommand{\divxi}{m_x}
\newcommand{\divyi}{m_y}
\newcommand{\divzi}{m_z}

\newcommand{\Aa}{{N_x}}
\newcommand{\AB}{{N_z}}
\newcommand{\Bb}{{N_y}}
\newcommand{\Aai}{i_x}
\newcommand{\ABi}{i_z}
\newcommand{\Bbi}{i_y}

\newcommand{\cA}[2]{\hat{\matA}_{#1}(#2)} 
\newcommand{\cB}[2]{\hat{\matB}_{#1}(#2)} 
\newcommand{\Ai}[1]{\matA_{#1}}           
\newcommand{\Aii}[2]{{\matA_{\tiny{#2}}^{(#1)}}}     
\newcommand{\Bi}[1]{\matB_{#1}}           
\newcommand{\Bii}[2]{{\matB_{\tiny{#2}}^{(#1)}}}     
\newcommand{\LY}{{L}}      
\newcommand{\ly}{l}      
\newcommand{\ND}{{N}}      
\newcommand{\nd}{n}      
\newcommand{\RT}{{R}}      

\newcommand{\rt}[1]{\RT_{#1}}

\newcommand{\DM}{{K}}      
\newcommand{\dm}[1]{\DM_{#1}}

\newcommand{\dmsummat}{\DM^{\text{h-mat}}_{\text{sum}}}
\newcommand{\dmsumpoly}{\DM^{\text{h-poly}}_{\text{sum}}}

\newcommand{\Xdimi}[1]{{\Aa}_{#1}}
\newcommand{\Zdimi}[1]{{\AB}_{#1}}
\newcommand{\Ydimi}[1]{{\Bb}_{#1}}

\usepackage{makeidx} 

\makeindex 
\begin{document}

\twocolumn[
\icmltitle{Cuboid Partitioning for Hierarchical Coded Matrix Multiplication}


\begin{icmlauthorlist}
\icmlauthor{Shahrzad Kiani}{t}
\icmlauthor{Nuwan Ferdinand}{t}
\icmlauthor{Stark C. Draper}{t}
\end{icmlauthorlist}

\icmlaffiliation{t}{Department of Electrical and Computer  Engineering, University of Toronto, Toronto, ON, Canada}
\icmlcorrespondingauthor{}{shahrzad.kianidehkordi@mail.utoronto.ca, \{nuwan.ferdinand, stark.draper\}@utoronto.ca}

\icmlkeywords{Machine Learning, icML}

\vskip 0.3in

\begin{abstract}
Coded matrix multiplication is a technique to enable straggler-resistant multiplication of large matrices in distributed computing systems. In this paper, we first present a conceptual framework to represent the division of work amongst processors in coded matrix multiplication as a cuboid partitioning problem. This framework allows us to unify existing methods and motivates new techniques. Building on this framework, we apply the idea of hierarchical coding~\cite{HIER:ISIT18} to coded matrix multiplication. The hierarchical scheme we develop is able to exploit the work completed by all processors (fast and slow), rather than ignoring the slow ones, even if the amount of work completed by stragglers is much less than that completed by the fastest workers. On Amazon EC2, we achieve a $37\%$ improvement in average finishing time compared to non-hierarchical schemes.

\end{abstract}
]	
\printAffiliationsAndNotice{\icmlEqualContribution}
\section{Introduction}
Large-scale matrix multiplication is a fundamental operation core to many data-intensive computational problems, including the training of deep neural networks. Such multiplication often cannot be performed in a single computer due to limited processing power and storage. Distributed matrix multiplication is necessary. While in an idealized setting highly parallelizable tasks can be accelerated proportional to the number of working nodes, in many cloud-based systems, slow working nodes, known as \emph{stragglers}, are a bottleneck that can prevent the realization of faster compute times~\cite{ Dean:2012}. Recent studies show that the effect of stragglers can be minimized through the use of error correction codes~\cite{SPEEDUP:TIT17, POLY:NIPS17, PRODUCT:ISIT17, MATDOT:2018, ENTGL:ISIT18}. This idea, termed {\em coded computation}, introduces redundant computations so that the completion of {\em any} fixed-cardinality subset of tasks suffices to realize the desired solution. A drawback of most methods of coded computing developed to date is that they rely only on the work completed by a set of the fastest workers, ignoring completely the work done by stragglers. Approaches to exploit the work completed by stragglers has been studied in~\cite{HIER:ISIT18}, which introduces idea of {\em hierarchical coding}.

In this paper, we establish an equivalence between task allocation in coded matrix multiplication and a geometric problem of partitioning a rectangular cuboid. The allocation of tasks in various prior coded matrix multiplication approaches -- Polynomial~\cite{POLY:NIPS17}, MATDOT~\cite{MATDOT:2018} codes, and others -- correspond to different partitions of the cuboid. Starting from this geometric perspective, we are able to extend the concept of hierarchical coding to any coded matrix multiplication. While the original concept of hierarchical coding~\cite{HIER:ISIT18} is introduced in the context of vector matrix multiplication using maximum distance separable (MDS) codes, its extension to general coded matrix multiplication is non-trivial. Cuboid partitioning visualization facilitates such extension.  

\section{Unifying geometric model}
\label{SEC:geometric}
Consider the problem of multiplying the two matrices $\inR{\matA}{\Aa}{\AB}$ and $\inR{\matB}{\AB}{\Bb}$ in a distributed system that consists of a \emph{master} node, and $\ND$ working nodes. We parallelize the computation of matrix product $\inR{\matA\matB}{\Aa}{\Bb}$ among the $\ND$ workers by providing each a subset of the data and asking each to carry out specific computations. We now present our conceptual framework wherein the decomposition of a matrix multiplication into smaller computations is visualized geometrically as the partitioning of a cuboid. 

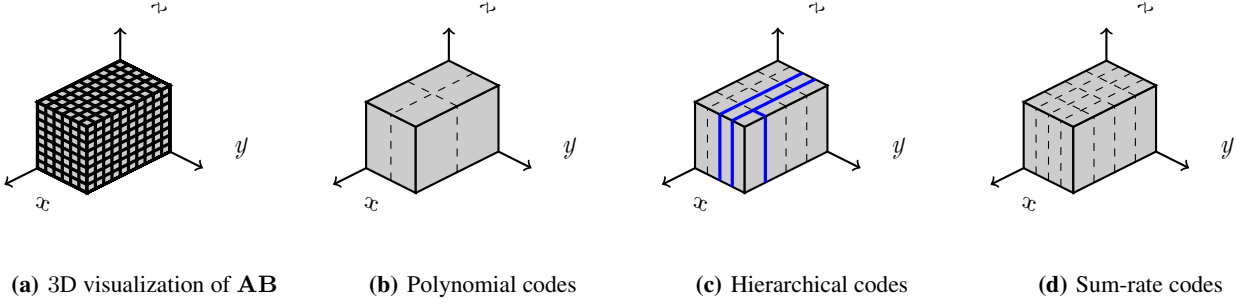
\begin{figure*}[h]
\centering 
	\subfigure[$3$D visualization of $\matA\matB$]{\tikzset{every mark/.append style={scale=1}}
	   \begin{tikzpicture} 
        [cube/.style={very thick,black},
            grid/.style={very thin,gray},
            axis/.style={->,black,thick},
            every node/.style={minimum size=1cm},on grid]

 \begin{scope}[scale=0.22,yshift=0,xshift=70]
 \begin{scope}[every node/.append style={yslant=-0.5},yslant=-0.5]
 [cube/.style={very thick,black},
            axis/.style={->,blue,thick}]
   \draw[axis] (5,5,0) -- (-4,-4,0) node[anchor=west]{$\Xdir$};
   \draw [step=0.5cm,very thick,fill=black!20!white](-2,-2) grid +(3,4) rectangle (-2,-2);
 \end{scope}

 \begin{scope}[every node/.append style={yslant=0.5},yslant=0.5]
 \draw[axis] (3,0,0) -- (8,-5,0) node[anchor=west]{$\Ydir$};

   \draw[step=0.5cm,very thick,fill=black!20!white] (1,-3) grid +(5,4) rectangle (1,-3);

 \end{scope}
 \begin{scope}[every node/.append style={
     yslant=0.5,xslant=-1},yslant=0.5,xslant=-1
   ]
   \draw[axis] (7,4,0) -- (9,6,0) node[anchor=west]{$\Zdir$};
   \draw[step=0.5cm,very thick,fill=black!20!white] (7,4) grid +(-5,-3)  rectangle (7,4);
 \end{scope}
 \end{scope}
 \end{tikzpicture}
         \label{FIG:dmm}}
	\quad	                   
	\subfigure[Polynomial codes]{\tikzset{every mark/.append style={scale=1}}
	   \begin{tikzpicture}
        [cube/.style={very thick,black},
            grid/.style={very thin,gray},
            axis/.style={->,black,thick},every node/.style={minimum size=1cm},on grid]

 \begin{scope}[scale=0.22,yshift=0,xshift=70]
 \begin{scope}[every node/.append style={yslant=-0.5},yslant=-0.5]
 [cube/.style={very thick,black},
            axis/.style={->,blue,thick}]
   \draw[axis] (5,5,0) -- (-4,-4,0) node[anchor=west]{$\Xdir$};
   \draw [step=0.5cm,thick,fill=black!20!white](-2,-2) rectangle +(3,4);
   \draw [dashed] (-0.5,-2) -- (-0.5,2);
 \end{scope}

 \begin{scope}[every node/.append style={yslant=0.5},yslant=0.5]
 \draw[axis] (3,0,0) -- (8,-5,0) node[anchor=west]{$\Ydir$};

   \draw[step=0.5cm,thick,fill=black!20!white] (1,-3) rectangle +(5,4);
   \draw [dashed] (3.5,-3) -- (3.5,1);
 \end{scope}

 \begin{scope}[every node/.append style={
     yslant=0.5,xslant=-1},yslant=0.5,xslant=-1
   ]
   \draw[axis] (7,4,0) -- (9,6,0) node[anchor=west]{$\Zdir$};
   \draw[step=0.5cm,thick,fill=black!20!white] (7,4) rectangle +(-5,-3);
   \draw [dashed] (7,2.5) -- (2,2.5);
   \draw [dashed] (4.5,4) -- (4.5,1);  
 \end{scope}
 \end{scope}
 \end{tikzpicture}         \label{FIG:POLY}}
	\quad       
	\subfigure[Hierarchical codes]{\tikzset{every mark/.append style={scale=1}}	
\begin{tikzpicture}
        [cube/.style={very thick,black},
            grid/.style={very thin,gray},
            axis/.style={->,black,thick},every node/.style={minimum size=1cm},on grid]

 \begin{scope}[scale=0.22,yshift=0,xshift=70]
 \begin{scope}[every node/.append style={yslant=-0.5},yslant=-0.5]
 [cube/.style={very thick,black},
            axis/.style={->,blue,thick}]
   \draw[axis] (5,5,0) -- (-4,-4,0) node[anchor=west]{$\Xdir$};
   \draw [step=0.5cm,thick,fill=black!20!white](-2,-2) rectangle +(3,4);
   \draw [dashed](-1.25,-2) -- (-1.25,2);
   \draw[blue,very thick] (0.25,-2) -- (0.25,2);
   \draw [blue,very thick](-0.5,-2) -- (-0.5,2);
 \end{scope}

 \begin{scope}[every node/.append style={yslant=0.5},yslant=0.5]
 \draw[axis] (3,0,0) -- (8,-5,0) node[anchor=west]{$\Ydir$};

   \draw[step=0.5cm,thick,fill=black!20!white] (1,-3) rectangle +(5,4);
   \draw [dashed] (4.75,-3) -- (4.75,1);
   \draw [dashed](3.5,-3) -- (3.5,1);
   \draw[blue,very thick] (2.25,-3) -- (2.25,1);

 \end{scope}

 \begin{scope}[every node/.append style={
     yslant=0.5,xslant=-1},yslant=0.5,xslant=-1
   ]
   \draw[axis] (7,4,0) -- (9,6,0) node[anchor=west]{$\Zdir$};
   \draw[step=0.5cm,thick,fill=black!20!white] (7,4) rectangle +(-5,-3);
   \draw [dashed](7,3.25) -- (2,3.25);
   \draw  [blue,very thick] (7,1.75) -- (2,1.75);
   \draw [blue,very thick](7,2.5) -- (2,2.5);
   
   \draw[ dashed] (5.75,4) -- (5.75,1);
   \draw[dashed] (4.5,4) -- (4.5,1);
   \draw [blue,very thick](3.25,1.75) -- (3.25,1);
   \draw [ dashed] (3.25,4) -- (3.25,1.75);
 \end{scope}
 \end{scope}
 \end{tikzpicture}     
       \label{FIG:HPOLY} }
       \quad   	
   \subfigure[Sum-rate codes]{\tikzset{every mark/.append style={scale=1}}	
\begin{tikzpicture}
        [cube/.style={very thick,black},
            grid/.style={very thin,gray},
            axis/.style={->,black,thick},every node/.style={minimum size=1cm},on grid]

 \begin{scope}[scale=0.22,yshift=0,xshift=70]

 \begin{scope}[every node/.append style={yslant=-0.5},yslant=-0.5]
 [cube/.style={very thick,black},
            axis/.style={->,blue,thick}]
   \draw[axis] (5,5,0) -- (-4,-4,0) node[anchor=west]{$\Xdir$};
   \draw [step=0.5cm,thick,fill=black!20!white](-2,-2) rectangle +(3,4);
   \draw [dashed](-1.25,-2) -- (-1.25,2);
   \draw[dashed] (0.25,-2) -- (0.25,2);
   \draw [dashed](-0.5,-2) -- (-0.5,2);
 \end{scope}

 \begin{scope}[every node/.append style={yslant=0.5},yslant=0.5]
 \draw[axis] (3,0,0) -- (8,-5,0) node[anchor=west]{$\Ydir$};
   \draw[step=0.5cm,thick,fill=black!20!white] (1,-3) rectangle +(5,4);
   \draw [dashed] (4.75,-3) -- (4.75,1);
   \draw [dashed](3.5,-3) -- (3.5,1);
   \draw[dashed] (2.25,-3) -- (2.25,1);
 \end{scope}

 \begin{scope}[every node/.append style={
     yslant=0.5,xslant=-1},yslant=0.5,xslant=-1
   ]
   \draw[axis] (7,4,0) -- (9,6,0) node[anchor=west]{$\Zdir$};
   \draw[step=0.5cm,thick,fill=black!20!white] (7,4) rectangle +(-5,-3);
   \draw [dashed](7,3.25) -- (2,3.25);
   \draw  [dashed] (7,1.75) -- (2,1.75);
   \draw [dashed](7,2.5) -- (2,2.5); 
   \draw[ dashed] (5.75,4) -- (5.75,1);
   \draw[dashed] (4.5,4) -- (4.5,1);
   \draw [dashed](3.25,1.75) -- (3.25,1);
   \draw [ dashed] (3.25,4) -- (3.25,1.75);
 \end{scope}
 
 \end{scope}
 \end{tikzpicture}        
      \label{FIG:BPOLY}} 
      	\quad         
       \caption{(a) $3$D visualization of the basic operations involved in matrix multiplication where $(\Aa,\AB,\Bb)=(10,8,6)$. Cuboid partitioning structure for (b) polynomial codes where $(\DM,\RT)=(4,4)$; (c) hierarchical codes where $\LY=4$ and $(\dm{i},\rt{i}) \in \{(8,8), (4,4), (3,3), (1,1)\}$; (d) sum-rate codes where $\LY=4$ and $(\DM_{\text{S-poly}},\RT_{\text{S-poly}})=(16,16)$.}
\end{figure*}

\subsection{$3$D model for matrix multiplication} 
\label{SEC:3dmm}
Standard techniques of matrix multiplication to compute the $\matA\matB$ product require $\Aa\AB\Bb$ \emph{\MAC}operations each of which is a multiply-and-accumulate. This \MAC operation $\fMAC: \mathbb R \times \mathbb R \times \mathbb R \rightarrow \mathbb R$ is defined pointwise as $\fMAC(a,b,c)=ab+c$. One method to compute each entry of $\matA\matB$ is iteratively to apply the \MAC operation $\AB$ times to compute an inner product. Each \MAC operation is indexed by a positive integer triple $(\Aai, \ABi, \Bbi) \in \idxset = [\Aa] \times [\AB] \times [\Bb]$\footnote{ $[\Aa] = \{1,\ldots,\Aa\}$ is the index set of cardinality $\Aa$.} such that the pairs $(\Aai, \ABi)$ and $(\ABi,\Bbi)$ index the entries of $\matA$ and $\matB$ that serve as the $a$ and $b$. In $3$D space, each integer triple $(\Aai, \ABi, \Bbi)$ can be considered as indexing a unit cube situated within a cuboid of integer edge lengths $(\Aa, \AB, \Bb)$, cf. Fig.~\ref{FIG:dmm}. The unit square in  the $\Xdir\Zdir$ or $ \Zdir \Ydir$ plane corresponding to index pair $(\Aai,\ABi)$ or $(\ABi,\Bbi)$ geometrically specifies the $\eA{\Aai}{\ABi}$ or $\eB{\ABi}{\Bbi}$ element in $\matA$ or $\matB$. Each unit square in the $\Xdir\Ydir$ plane represents an entry of $\matA\matB$.

\subsection{$3$D model for coded matrix multiplication}
\label{SEC:3dcoded}
We now derive previously presented coded schemes from the cuboid partitioning perspective. We recall the terminology and setup introduced in the previous literature~\cite{SPEEDUP:TIT17}. In coded matrix multiplication, the $\matA\matB$ product is first partitioned into $\DM$ equal-sized computations. The master then encodes the data involved in each of the $\DM$ computations to yield a larger set of $\ND$ {\em encoded} tasks. Each task is given to a distinct worker. The master can recover the original $\DM$ computations by decoding any $\RT$ completed tasks from any set of workers. We term $\DM$ the {\em information dimension} and $\RT$ the {\em recovery threshold}.

To partition the overall computation of the $\matA\matB$ product into $\DM$ equal-sized computations, the master first partitions the data. Once the data $\matA$ and $\matB$ are partitioned, certain pairs of submatrices can be matched up to yield the $\DM$ computations. The partitioning of the $\matA\matB$ product into the $\DM$ computations can be visualized as a partitioning of the cuboid. The $\DM$ distinct computations are represented by $\DM$ equal-sized subcuboid partitions. We use \emph{\infoblock} to refer to such subcuboids. In the following, we first introduce the partitioning structure of polynomial codes~\cite{POLY:NIPS17}; we then generalize this idea to all previous coding schemes.

\textbf{Polynomial codes:} One way to achieve information dimension $\DM$ in polynomial codes is to divide $\matA$ and $\matB$ into $\sqrt{\DM}$ matrices (for simplicity we assume $\sqrt{\DM}$ is an integer). Partition $\matA$ horizontally into $\sqrt{\DM}$ matrices\footnote{$\Ai{\seti{1} \times \seti{2}}$ is a collection of $\seti{1}$ rows and $\seti{2}$ columns of $\matA$.} $\Ai{\setxi{i} \times [\AB]}$, where $i \in [\sqrt{\DM}]$ and $ \setxi{i} = [\floor{{\Aa}/{\sqrt{\DM}}}]+ (i-1)\floor{{\Aa}/{\sqrt{\DM}}}$. Partition $\matB$ vertically into $\sqrt{\DM}$ matrices $\Bi{[\AB]\times \setyi{i}}$, where $i \in [\sqrt{\DM}]$ and $ \setyi{i}=[\floor{{\Bb}/{\sqrt{\DM}}}]+(i-1)\floor{{\Bb}/{\sqrt{\DM}}}$. Each computation is the matrix product $\Ai{\setxi{i} \times [\AB]}\Bi{[\AB]\times \setyi{j}}$, where $i,j \in [\sqrt{\DM}]$. This decomposition {\em slices} the cuboid into $\DM$ equal-sized subcuboids by making $\sqrt{\DM}-1$ parallel cuts along the $\Xdir$-axis and $\sqrt{\DM}-1$ parallel cuts along the $\Ydir$-axis, cf. Fig.~\ref{FIG:POLY} for $\DM=4$.


\textbf{Generalized coded matrix multiplication}: In general, all possible cuboid partitions that arise in coded matrix multiplication can be clustered into eight possible categories. Each category is defined by cutting the cuboid along a specific subset of directions $\{\Xdir,\Ydir,\Zdir \}$. For example, product codes~\cite{ PRODUCT:ISIT17} and polynomial codes~\cite{POLY:NIPS17} slice along the $\Xdir$- and $\Ydir$- axes. MATDOT codes~\cite{MATDOT:2018} slice along the  $\Zdir$- axis. Generalized polyDOT codes~\cite{MATDOT:2018} and entangled polynomial codes~\cite{ENTGL:ISIT18} slice the cuboid along all axes.

\section{Hierarchical matrix multiplication}
\label{SEC:HMM}
We now employ our geometric insight to design the general hierarchical coded matrix multiplication in three phases.



\textbf{Data and cuboid partitioning phase:} In contrast to the one-phase cuboid partitioning of coded matrix multiplication scheme, in a hierarchical scheme the master partitions the cuboid in two steps. It first divides the cuboid into $\LY$ subcuboids each of which we think of as a \emph{\layer} of computation. The term \emph{\taskbox} is used for such subcuboids partitions in the first step of partitioning.  It then partitions the $\ly$th \taskbox into $\dm{\ly}$ equal-sized \infoblocksno. 

The $\ly$th \taskbox is described by the set $\seti{\ly} = \setxi{\ly} \times \setzi{\ly} \times \setyi{\ly}$ where $\setxi{\ly}, \setzi{\ly}$ and $\setyi{\ly}$ are, respectively, subsets of (generally) consecutive elements of $[\Aa],[\AB]$ and $[\Bb]$. Such \taskbox corresponds to a set of \MAC operations indexed by $(\Aai,\ABi,\Bbi) \in \seti{\ly}$. The $\ly$th \taskbox can be visualized as a cuboid of dimensions $\Xdimi{\ly} \times \Ydimi{\ly} \times \Zdimi{\ly}$, where $\Xdimi{\ly}=\abs{\setxi{\ly}}, \Zdimi{\ly}=\abs{\setzi{\ly}}$ and $ \Ydimi{\ly} = \abs{\setyi{\ly}}$. The master starts by grouping $\matA$ into $\LY$ \taskmtxs $\{\Ai{\setxi{\ly} \times\setzi{\ly}}|\, \ly \in [\LY]\}$ and grouping $\matB$ into $\LY$ \taskmtxs $\{\Bi{\setzi{\ly} \times\setyi{\ly}}|\, \ly \in [\LY]\}$ such that the $\matA\matB$ product is decomposed into $\LY$ computations $\Ai{\setxi{\ly} \times\setzi{\ly}}\Bi{\setzi{\ly} \times\setyi{\ly}}$. To denote the $i$th element of $\setxi{\ly}$ we write $\setxij{\ly}{i}$, which is a row-index into $\matA$. Similarly, $\setzij{\ly}{i}$ and $\setyij{\ly}{i}$ are column-indices into $\matA$ and $\matB$, respectively.

In the second step of partitioning, the master subdivides $\Ai{\setxi{\ly}\times\setzi{\ly}}$ into $\divx \divz$ equal-sized \datachunks denoted as $\Aii{\ly}{\divxi,\divzi}$, where $(\divxi,\divzi) \in [\divx] \times [\divz]$ and $\ly \in [\LY]$. The \datachunk $\Aii{\ly}{\divxi,\divzi}$ contains all elements $(\matA)_{\setxij{\ly}{\Aai},\setzij{\ly}{\ABi}}$, where $ \Aai \in  [{\Xdimi{\ly}}/{\divx}] + (\divxi-1){\Xdimi{\ly}}/{\divx} $ and $ \ABi \in  [{\Zdimi{\ly}}/{\divz}] + (\divzi-1){\Zdimi{\ly}}/{\divz}$. Likewise, the master subdivides $\Bi{\setzi{\ly}\times\setyi{\ly}}$ into $\divz \divy$ equal-sized \datachunksno: $\Bii{\ly}{\divzi,\divyi}$, where $(\divzi,\divyi) \in [\divz] \times [\divy]$.  The \datachunk $\Bii{\ly}{\divzi,\divyi}$ contains all elements $(\matB)_{\setzij{\ly}{\ABi},\setyij{\ly}{\Bbi}}$ where $ \ABi \in  [{\Zdimi{\ly}}/{\divz}] + (\divzi-1){\Zdimi{\ly}}/{\divz} $ and $\Bbi \in  [{\Ydimi{\ly}}/{\divy}] + (\divyi-1){\Ydimi{\ly}}/{\divy}$. The partitioning of $\Ai{\setxi{\ly}\times \setzi{\ly}}$ and $\Bi{\setzi{\ly}\times \setyi{\ly}}$ divides the $\ly$th \taskbox into equal-sized \infoblocks each of dimensions ${\Xdimi{\ly}}/{\divx} \times {\Zdimi{\ly}}/{\divz} \times {\Ydimi{\ly}}/{\divy}$. The information dimension used in the $\ly$th layer is equal to the number of \infoblocks in that layer, \ie $\dm{\ly}=\divx\divz\divy$. 
In Fig.~\ref{FIG:HPOLY}, the decomposition of $\matA\matB$ into \layers of computation is depicted by the bold solid (blue) lines whereas \taskboxes are partitioned into \infoblocks by dashed lines. 



\textbf{Data encoding and distribution phase:} The master generates $\ND$ pairs of \encdatachunks $\{(\cA{\ly}{\nd},\cB{\ly}{\nd})| \, \nd \in [\ND]\}$ from $2\dm{\ly}$ \datachunks $\{\Aii{\ly}{\divxi,\divzi},\Bii{\ly}{\divzi,\divyi} | \, \divxi \in [\divx], \, \divzi \in [\divz], \, \divyi \in [\divy] \}$. For instance, in~\cite{POLY:NIPS17} the polynomials used to encode the \datachunks of layer $\ly$ are $\cA{\ly}{x} = \sum_{\divxi} \Aii{\ly}{\divxi,1} x^{\divxi-1}$ and $\cB{\ly}{x} = \sum_{\divyi} \Bii{\ly}{1,\divyi} x^{(\divyi-1)\divx}$. 

\textbf{Worker computation and decoding phase:} The $\nd$th worker sequentially computes its $\LY$ jobs, $\cA{1}{\nd}\cB{1}{\nd}$ through $\cA{\LY}{\nd}\cB{\LY}{\nd}$, sending completed jobs to the master as soon as they are finished. 
To recover all the \infoblocks that make up the $\ly$th \layer of computation, the master must receive at least $\rt{\ly}$ jobs from the $\ND$ workers, i.e., a subset of size at least $\rt{\ly}$ of $\{\cA{\ly}{\nd}\cB{\ly}{\nd} \, | \, \nd \in [\ND]\}$. The master then decodes the output of each layer using a decoding algorithm. For instance, if the $\ly$th layer is encoded by polynomial codes, the master can use a Reed-Solomon decoder~\cite{REED:2009}.



\section{Evaluation}
\begin{figure}
\centering  
		\begin{tikzpicture}[scale=0.67]
	\begin{axis}[
	height=10cm,
	width=10cm,
	grid=major,
	xlabel={Number of layers $L$},
	ylabel={Average finishing time (sec)},
legend style={at={(0.45,0.83)},anchor=west,nodes=right},	
	axis on top,xmin=1, xmax=12, ymin=0.8, ymax=2.2]

	\addlegendentry{Uncoded}
		\addplot [line width=0.5mm, color=black, dashdotted, every mark/.append style={solid, fill=black},mark=triangle*] coordinates {
 	(1,2.17)
 	(2,2.17)
 	(4,2.17)
 	(6,2.17)
 	(8,2.17)
 	(10,2.17)
 	(12,2.17)
	};

	\addlegendentry{Poly, $\DM_{\text{poly}}=11$}
		\addplot [line width=0.5mm, color=black, dashed, every mark/.append style={solid, fill=black},mark=square*] coordinates {
 	(1,    1.8728)
 	(2,    1.8728)
 	(4,    1.8728)
 	(6,    1.8728)
 	(8,    1.8728)
 	(10,    1.8728)
 	(12,    1.8728)
	};	
	
	\addlegendentry{H-poly, $\sum_{\ly} \dm{\ly}=11\LY$}
		\addplot [line width=0.5mm, color=blue, solid, every mark/.append style={solid, fill=blue},mark=*] coordinates {
 	(1,1.8728)
 	(2,1.4161)
 	(4,1.2944)
 	(6,1.2076)
 	(8,1.1823)
 	(10,1.1728)
 	(12,1.1688)
	};	

		\addlegendentry{S-poly, $\DM_{\text{S-poly}} =11\LY$}
		\addplot [line width=0.5mm, color=red, dotted, every mark/.append style={solid, fill=red},mark=otimes*] coordinates {
 	(1,1.8728)
 	(2,1.2544)
 	(4,1.1104)
 	(6,1.0364)
 	(8,1.0174)
 	(10,1.0025)
 	(12,0.9854) 	
	};



	\end{axis}
	\end{tikzpicture} 
\caption{The average finishing time vs. $\LY$, where $(\Aa,\AB,\Bb)$ $=(10000,1000,3000)$ and $\DM_{\text{poly}} = \DM_{\text{S-poly}}/L = \DM_{\text{sum}}/L=11$.}  	          
\label{EC2} \vspace*{-3.5ex} 
\end{figure}
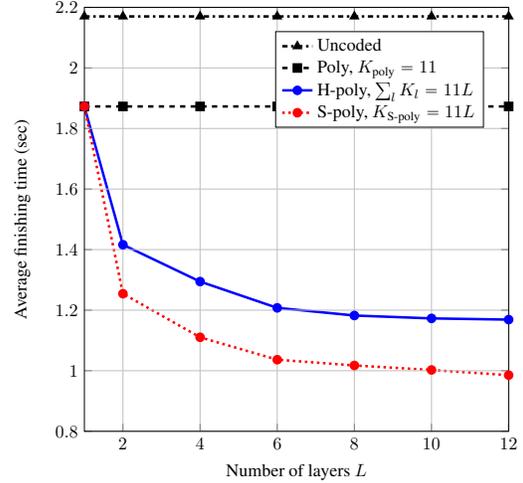

In Amazon EC2, we implement a large matrix multiplication $\matA\matB$ on $\ND=16$ t2.micro instances in parallel using the mpi4py library. In Fig.~\ref{EC2} we plot the average finishing time which counts only the time of computation vs. number of layers $\LY$. Since in EC2 we rarely observe stragglers in small-scale distributed system ($\ND=16$), we artificially assign workers to be straggler with probability $0.5$. Workers that are designated stragglers are tasked by one more extra computation than non-stragglers per layer. We measure the average finishing time (over 10 instances) for the uncoded scheme, polynomial, hierarchical polynomial, and sum-rate~\cite{EXPLOIT:ISIT18} polynomial codes. While sum-rate polynomial coding is an alternate design to exploit stragglers, it has a cuboid partitioning structure similar to that of its hierarchical polynomial coding equivalent, cf. Fig.~\ref{FIG:BPOLY}. The distinction between hierarchical and sum-rate polynomial codes is that for the sum-rate scheme the data involved in the information blocks are used to generate a single polynomial code (instead of $\LY$). The rate of the sum-rate scheme, $\DM_{\text{S-poly}}/\ND$, is equal to the sum of the per-layer rates used in the hierarchical scheme, \ie $\DM_{\text{S-poly}}/\ND=\sum_{\ly}\dm{\ly}/\ND$. In uncoded scheme, the computation is split evenly amongst processors (each $1/\ND$) without any redundancy. Fig.~\ref{EC2} shows that the hierarchical scheme with $\LY=12$ achieves $37\%$ and $46\%$ improvements in comparison to the polynomial coding and uncoded scheme, respectively. While the average finishing times of sum-rate scheme lower bounds the average finishing time of hierarchical scheme, in the sum-rate coding the master deals with decoding a polynomial code of rate $\DM_{\text{S-poly}}/\ND$ which is much more computationally intensive than the decoding of the rate $\{\dm{\ly}/\ND\}_{\{\ly \in [\LY]\}}$ polynomial codes used in the hierarchical scheme.

\section{Conclusion}
In this paper we connect the task allocation problem that underlies coded matrix multiplication to a geometric question of cuboid partitioning. Through this geometric perspective we introduce hierarchical coded matrix multiplication. On Amazon EC2, we show a $37\%$ improvement in the average finishing time when compared to non-hierarchical schemes.

\clearpage
\bibliographystyle{icml2019}
\bibliography{reference}
\clearpage

\end{document}